\documentclass{article}

\usepackage[preprint, nonatbib]{neurips_2022}

\usepackage[utf8]{inputenc} 
\usepackage[T1]{fontenc}    
\usepackage{hyperref}       
\usepackage{url}            
\usepackage{booktabs}       
\usepackage{amsfonts}       
\usepackage{nicefrac}       
\usepackage{microtype}      
\usepackage{xcolor}         
\usepackage{bm}
\usepackage{amsmath}
\usepackage{graphicx}
\usepackage{array}
\usepackage{caption}
\usepackage{subcaption}

\title{Supplementing Recurrent Neural Networks with Annealing to Solve Combinatorial Optimization Problems}

\author{%
  Shoummo Ahsan Khandoker \\
  Department of Computer Science\\
  BRAC University \\
  \texttt{shoummo.ahsan.khandoker@g.bracu.ac.bd} \\
  \And
  Jawaril Munshad Abedin \\
  Department of Computer Science\\
  BRAC University \\
  \texttt{jawaril.munshad.abedin@g.bracu.ac.bd} \\
  \AND
  Mohamed Hibat-Allah \\
  Department of Physics and Astronomy \\
  University of Waterloo \\
  Vector Institute for Artificial Intelligence \\
  \texttt{mohamed.hibat.allah@uwaterloo.ca} \\
}

\begin{document}

\maketitle

\begin{abstract}
    Combinatorial optimization problems can be solved by heuristic algorithms such as simulated annealing (SA) which aims to find the optimal solution within a large search space through thermal fluctuations. The algorithm generates new solutions through Markov-chain Monte Carlo techniques. This sampling scheme can result in severe limitations, such as slow convergence and a tendency to stay within the same local search space at small temperatures. To overcome these shortcomings, we use the variational classical annealing (VCA) framework~\cite{hibat2021variational} that combines autoregressive recurrent neural networks (RNNs) with traditional annealing to sample solutions that are uncorrelated. In this paper, we demonstrate the potential of using VCA as an approach to solving real-world optimization problems. We explore VCA's performance in comparison with SA at solving three popular optimization problems: the maximum cut problem (Max-Cut), the nurse scheduling problem (NSP), and the traveling salesman problem (TSP). For all three problems, we find that VCA outperforms SA on average in the asymptotic limit by one or more orders of magnitude in terms of relative error. Interestingly, we reach large system sizes of up to $256$ cities for the TSP. We also conclude that in the best case scenario, VCA can serve as a great alternative when SA fails to find the optimal solution.
\end{abstract}
\section{Introduction}\label{sec:intro}

Combinatorial optimization is widely used in many areas of science such as physics, computer science, and biology. It also has a wide range of applications in the industry including without limitation to supply chain, energy, and transportation. Providing an efficient solution to combinatorial optimization problems can boost scientific progress and provide optimal solutions to lots of industry problems. Nevertheless, a large class of these problems is NP-Hard~\cite{Lucas2014}. Thus, devising clever algorithms is needed to cope with the exponential growth in the search space that contains all possible combinatorial solutions. In particular, a heuristic algorithm based on the concept of annealing was devised and is known as simulated annealing (SA) or classical annealing (CA)~\cite{SApaper}. The latter is inspired by metallurgy, where slowly reducing the temperature of a metal can result in a more stable configuration with a lower energy~\cite{SApaper}. Given the Markovian nature of these algorithms, they can result in getting stuck in a local minima due to the rugged nature of some optimization landscapes. 

In this paper, we aim to use tools from machine learning for the purpose of improving traditional implementations of SA. In particular, this work is based on the findings of Ref.~\cite{hibat2021variational} that uses recurrent neural networks (RNNs) to variationally simulate the classical version of annealing within a framework called variational classical annealing (VCA)~\cite{hibat2021variational}. Here, we apply RNNs through VCA to find the optimal solution to real-world optimization problems namely the maximum cut problem (Max-Cut), the nurse scheduling problem (NSP), and the traveling salesman problem (TSP). This work has shown that this variational implementation is capable of outperforming SA on average and in the asymptotic limit of slow annealing. We note that the use of RNNs in our work allows us to avoid the expensive use of GPU memory resources, and to reach system sizes beyond $100$ cities for the TSP. This limitation can be found in traditional Transformer models~\cite{bresson2021transformer}.

\section{Related Work}

Similar studies have proposed the use of a parameterized ansatz to approximate the ground state of combinatorial optimization problems. In particular, Ref.~\cite{gomes2019classical} introduces a technique called classical quantum optimization (CQO) where they use neural-network quantum states (NQS) \cite{carleo2017solving} to find the ground state of optimization problems that have been cast into a cost Hamiltonian (also called the energy function)~\cite{gomes2019classical,Sinchenko2019,zhao2020natural}. The NQS ansatz aims to minimize the cost Hamiltonian by training its parameters using the stochastic reconfiguration (or natural gradient) optimization technique \cite{sorella1998green}. Our study is based on the findings of Ref.~\cite{hibat2021variational} that adds the annealing component on top of CQO. This study was further extended to the task of finding the ground state of quantum systems with frustration~\cite{RNNAnnealing}. 

Furthermore, there have been studies that leverage the Transformer architecture~\cite{vaswani2017attention} to parameterize the solution distribution of the TSP~\cite{bresson2021transformer, kool2018attention, deudon2018learning}. The aforementioned papers encode the TSP cities using a transformer encoder but take different approaches to decoding the next city in the partial (uncomplete) tour. Ref.~\cite{kool2018attention} queries the first and last cities in the partial tour, and a combined representation of all the cities to decode the next city. For training, the authors use reinforcement learning (RL) with a greedy baseline. Ref.~\cite{deudon2018learning} decodes the next city with a query consisting of the last 3 cities in the partial tour, and the training being done with the Actor-Critic RL~\cite{deudon2018learning}. Afterward, the solution generated by the transformer architecture is further improved using the 2-OPT heuristic. Ref.~\cite{bresson2021transformer} decodes the next city autoregressively, meaning the query is constructed using all the cities in the partial tour. The training is performed in an RL framework with a greedy baseline similar to Ref.~\cite{kool2018attention}.

Additionally, thermal fluctuations have been supplemented to the process of training binary neural networks within an approach known as the Bayesian learning rule~\cite{bayeslearningrule2020}. Thermal fluctuations and entropy have also been used to improve Actor-Critic RL under a framework called Soft Actor-Critic (SAC) RL~\cite{SAC2018}. We also see the use of annealing in machine learning being put forward in the training process of variational auto-encoders~\cite{VAEannealing}. Similarly to the principle of annealing, the gradual increase in problem complexity has been explored to solve optimization problems under a framework called curriculum learning~\cite{curriculumlearning}. We finally note that a summary of the progress made by the machine learning community for the task of solving optimization problems can be found in Ref.~\cite{BENGIO2021405}.

\section{Variational Classical Annealing}\label{sec:vca}

For an energy function (or a Hamiltonian) $H(\boldsymbol{X})$ that encodes an optimization problem, our aim is to find the ground state $\boldsymbol{X}^*$ by optimizing a set of a model's parameters $\boldsymbol{\lambda}$. The variational classical annealing (VCA) algorithm introduced in Ref.~\cite{hibat2021variational}, aims to produce a probability distribution $P_{\boldsymbol{\lambda}}$ that approximates the Boltzmann distribution at zero temperature through annealing and optimization of the parameters $\boldsymbol{\lambda}$. The peak of this Boltzmann distribution provides the optimal solutions to $H(\boldsymbol{X})$. This algorithm is inspired by the use of variational autoregressive networks to solve statistical mechanics systems \cite{wu2019solving}.  

Autoregressive models originally aim to use the observation in the previous time step as input to the subsequent time step as it usually done in natural language processing. In the context of our work, we harness the power of autoregressive models for the purposes of solving optimization problems by sequentially sampling solutions to these problems~\cite{hibat2021variational}. This sampling scheme allows to obtain perfect samples as opposed to Metropolis sampling, which can produce correlated samples, and which can get stuck in local minima if the optimization problem has a rugged optimization landscape~\cite{condmat7020038}.

Within the setting of VCA, we use RNNs as our model to be optimized. This scheme is performed by minimizing the variational free energy function
\begin{equation}
    F_{\boldsymbol{\lambda}}(t) = \langle H(\boldsymbol{X}) \rangle_{\boldsymbol{\lambda}} - T(t)S(P_{\boldsymbol{\lambda}}).
\end{equation}
The gradient descent learning on this cost function is used to update $\boldsymbol{\lambda}$ at every annealing stage. Here, $\langle H(\boldsymbol{X}) \rangle_{\boldsymbol{\lambda}}$ is the average Hamiltonian over the RNN distribution $P_{\boldsymbol{\lambda}}$,  $T(t)$ denotes the temperature at time $t$, and $S(P_{\boldsymbol{\lambda}})$ is the Shannon entropy. The latter is formulated as
\begin{equation}
    S(P_{\boldsymbol{\lambda}}) = -\sum_{i}P_{\boldsymbol{\lambda}}(\boldsymbol{X}_i)\log(P_{\boldsymbol{\lambda}}(\boldsymbol{X}_i)),
\end{equation}
where the sum iterates over all the states $\{\boldsymbol{X}_i\}$ in the search space. Computing the true entropy by considering all the states in the exact state space is intractable in general. However, one can approximate the entropy by taking a statistical average over the autoregressive samples obtained from the distribution $P_{\boldsymbol{\lambda}}$ as shown in Appendix.~\ref{sec:hyperparams}. The VCA framework is based on temperature annealing. The latter was found to be helpful in avoiding mode collapse~\cite{wu2019solving} and in an adiabatic evolution to the lowest energy~\cite{hibat2021variational}. In our annealing scheme, we use a linear annealing schedule characterized by $T(t)=T_{0}(1-t)$ where time $t\in[0,1]$ and $T_0$ is the initial temperature. 

VCA initially performs a series of warm-up steps for the purpose of bringing a randomly initialized RNN probability distribution to the minimum of the free energy $F_{\boldsymbol{\lambda}}(0)$ by performing $N_{\text{warmup}}$ gradient descent steps. Throughout this initial equilibrating process, the temperature is kept constant at $T(0)=T_0$. Once the warm-up steps are completed, the annealing process begins. Any time the temperature is cooled from $T(t')$ to $T(t'')$, the shape of the free energy landscape of the system changes depending on the magnitude of $\Delta t = t'' - t'$. To equilibrate the RNN to a minimum of the new free energy $F_{\boldsymbol{\lambda}}(t'')$, we perform $N_\text{train}$ gradient descent steps. This combination of annealing and re-equilibration is performed until $t = 1$ where $T(1)=0$, marking the end of the annealing process. At this point, the probability distribution $P_{\boldsymbol{\lambda}}$ is expected to be maximized around states with an energy equal or close to the ground state energy $H(\boldsymbol{X}^*)$. 

In our implementation, we parameterize time as $t=i/N_{\text{annealing}}$, where $N_{\text{annealing}}$ is a user-defined hyperparameter denoting the number of annealing steps which effectively determines the speed of the annealing process, such that $i$ iterates over $0, \dots, N_{\text{annealing}}$.

\section{Optimization Problems}\label{sec:op}

\subsection{The Nurse Scheduling Problem (NSP)}\label{sec:nsp}

We now focus our attention on the optimization problems we tackle using the VCA algorithm. The first one is NSP which belongs to the class of scheduling problems in the field of operations research, and it is known to be NP-hard \cite{osogami2000classification}. NSP aims to assign nurses to specific shifts in a hospital under a set of imposed constraints. The imposed constraints make it hard to find satisfactory solutions to the problem. In this work, we apply VCA to the formulation of NSP introduced in Ref. \cite{ikeda2019application}. 

In this formulation, we have the following constraints: the hard nurse constraint,  the hard shift constraint, and the soft nurse constraint. The hardness determines the importance of respecting the constraint. The hard nurse constraint requires that no nurse works for two consecutive shifts as they require sufficient rest after a shift. The hard shift constraint emphasizes the need to deploy enough nurses to handle a given shift. Finally, the soft nurse constraint favors solutions with an even distribution of nurses assigned to shifts. In our paper, the terms 'shift' and 'day' are synonymous.

To better understand NSP, we take a look at what defines such a problem. In a hospital, we have $N$ individual nurses given by $n \in \{1,\dots,N\}$ and $D$ working days given by $d \in \{1,\dots,D\}$. A solution to the NSP problem could be represented by a matrix
\begin{equation*}
    \boldsymbol{X_M} = 
    \begin{bmatrix}
    x_{1,1} & \dots & x_{1,D} \\
    \vdots & \ddots & \vdots \\
    x_{N,1} & \dots & x_{N,D} \\
    \end{bmatrix},
\end{equation*}
where $\boldsymbol{X_M} \in \{0,1\}^{N \times D}$. If nurse $n$ has been assigned to day $d$, then $x_{n,d} = 1$, otherwise $x_{n,d} = 0$. We use the flattened vector representation of $\boldsymbol{X_M}$ to represent our solution, denoted by $\boldsymbol{X} \in \{0,1\}^{ND}$. With index manipulation, the matrix elements of $\boldsymbol{X_M}$ map to the elements in $\boldsymbol{X} = [x_{m(1,1)}, \dots, x_{m(N,D)}]$ using $m(n,d)=D(n-1)+d$.

The three constraints are formulated individually into separate quadratic penalty functions \cite{ikeda2019application}. The hard nurse constraint, which is quantified by
\begin{equation}\label{eq:nsp_1}
    H_1(\boldsymbol{X}) = \sum_{n = 1}^{N}\sum_{d = 1}^{D} \alpha x_{m(n,d)}x_{m(n,d+1)},
\end{equation}
penalizes a solution that has instances of a nurse $n$ working two days in a row. Every time this violation occurs, a penalty of $\alpha$ is added.

The hard shift constraint has been formulated using the workforce required on a particular day $W(d)$. The constraint Hamiltonian
\begin{equation}\label{eq:nsp_2}
    H_2(\boldsymbol{X}) = \sum_{d=1}^{D}\left( \sum_{n = 1}^{N} x_{m(n,d)} - W(d) \right)^2,
\end{equation}
aims to equalize the accumulated contribution of the nurses assigned to a working day and the actual amount of workforce required on that day. A penalty is incurred when there are not enough nurses assigned to a day, or conversely, when there is a surplus of nurses.

Lastly, we have the soft nurse constraint which promotes equal distribution of all nurses across the working days. This constraint is given as follows:
\begin{equation}\label{eq:nsp_3}
    H_3(\boldsymbol{X}) = \sum_{n = 1}^{N}\left( \sum_{d = 1}^{D}x_{m(n,d)} - F(n) \right)^2.
\end{equation}
The soft constraint term $H_3(\boldsymbol{X})$ has been formulated using $F(n)$ that gives us the number of days a nurse $n$ wishes to work. If equal distribution of nurses is a desirable outcome, then setting at least $F(n) = \lfloor D/N \rfloor$ for all nurses is required as this ensures a fair workload when $D$ is a multiple of $N$. 

We can sum the three penalty terms shown in Eq.~\eqref{eq:nsp_1}, \eqref{eq:nsp_2}, \eqref{eq:nsp_3} to get the resultant energy function
\begin{equation}\label{eq:nsp_4}
    H(\boldsymbol{X}) = H_1(\boldsymbol{X}) + \lambda H_2(\boldsymbol{X}) + \gamma H_3(\boldsymbol{X}),
\end{equation}
where $\lambda$ and $\gamma$ are real coefficients that scale the penalty terms $H_2(\boldsymbol{X})$ and $H_3(\boldsymbol{X})$ respectively. $H(\boldsymbol{X})$ is minimized when the least number of constraints are broken. If all the constraints are obeyed then the optimal energy $H(\boldsymbol{X^{*}})=0$. However, it should be noted that for certain combinations of $N$ and $D$, it is not always possible for the ground state $\boldsymbol{X}^{*}$ to respect all the constraints, resulting in $H(\boldsymbol{X^{*}}) > 0$.

Interestingly, given the quadratic form of $H(\boldsymbol{X})$ the binary nature of $\boldsymbol{X}$, $H(\boldsymbol{X})$ can be represented as a quadratic unconstrained binary optimization (QUBO) model~\cite{ikeda2019application, glover2019quantum} such that
\begin{equation*}
    H(\boldsymbol{X}) = \sum_{i = 1}^{ND}\sum_{j= i}^{ND} q_{ij}x_{i}x_{j} + c,
\end{equation*}
where coefficient $q_{i,j}$ and constant $c$ are derived from Eq.~\eqref{eq:nsp_4}. This expression can be made more compact as
\begin{equation*}
    H(\boldsymbol{X}) = \boldsymbol{X}^T\boldsymbol{QX} + c,
\end{equation*}
such that $\boldsymbol{Q}\in \mathbb{R}^{ND\times ND}$ is a square matrix with matrix elements given by $q_{ij}$.

\subsection{The Maximum Cut Problem (Max-Cut)}

The second problem we tackle with VCA is the Max-Cut problem that has been known to be at least NP-hard \cite{karp1972reducibility}. The Max-Cut problem is defined as follows: Given an undirected graph $G(V, E)$, we make a cut along the edges of $G$ to get two complementary sets of vertices such that the number of edges between the two sets is maximized. In other words, if $E_{\text{cut}} \subset E$ is the set of edges bridging the two complementary sets of vertices, we wish to maximize its size $|E_{\text{cut}}|$. In the context of our study, we work with unweighted graphs.

To model the partition across the graph, we can set a value of either $1$ or $0$ to the vertices to label the partitions where they belongs. Therefore, any solution to a graph of $N$ vertices is given by $\boldsymbol{X}=(x_1, \dots, x_N)$ where $x_i \in \{0,1\}$. This mapping allows us to use an energy function $H(\boldsymbol{X})$ that computes the negative of the sum of the number of edges belonging in $E_{\text{cut}}$. This step is done by summing the following expression over all the edges in set $E$:
\begin{equation}\label{eq:mc}
    H(\boldsymbol{X})=-\sum_{(i,j)\in E}(x_i+x_j-2x_ix_j).
\end{equation}
Taking the negative converts Max-Cut into a minimization problem. An edge connecting $x_i$ and $x_j$ exists in $E_{\text{cut}}$ when $x_i \neq x_j$. The latter provides a contribution of $-1$ to $H(\boldsymbol{X})$. For simplicity, we note that the expression of $H(\boldsymbol{X})$ is equivalent to the following Kronecker Delta expression
\begin{equation*}
    H(\boldsymbol{X})=-\sum_{(i,j)\in E}1-\delta_{x_ix_j}.
\end{equation*}

Note that $H(\boldsymbol{X})$ in Eq. (\ref{eq:mc}) is quadratic due to the binary nature of variables $x_i$. Therefore, $H(\boldsymbol{X})$ can be cast into a QUBO form as follows
\begin{equation}\label{eq:mc_qubo}
    H(\boldsymbol{X})=\boldsymbol{X}^T\boldsymbol{QX},
\end{equation}
where $\boldsymbol{Q}\in \mathbb{R}^{N\times N}$ is the QUBO square matrix with matrix elements derived from Eq. (\ref{eq:mc}).

\subsection{The Traveling Salesman Problem (TSP)}

TSP is another famous combinatorial optimization problem that we attempt to solve with VCA. The aim of TSP is to visit a set of cities exactly once and return to the starting city in the tour with a specific order that minimizes the total cost of travel. In our study, the cost is the total distance traveled. We are particularly interested in the 2D Euclidean TSP which is known to be NP-complete \cite{papadimitriou1977euclidean}.

Given $N$ cities on a 2D Cartesian plane, a solution tour looks as $\boldsymbol{X}=(\pi(1), \dots, \pi(N))$ where $\pi$ is a permutation of $N$ cities. Each city $\pi(i)$ has a tuple of $x$ and $y$ coordinates $(x_{\pi(i)},y_{\pi(i)})$. An important property of a valid solution that should be noted is that every city on the tour should be unique. We also remark that the permutation order of cities is translation invariant. The latter property is exploited in the construction of our parameterized model. The energy function of the 2D Euclidean TSP is given by the sum of the Euclidean distances between consecutive cities, including the trip from the last city back to the first as follows
\begin{equation}\label{eq:tsp_6}
    H(\boldsymbol{X})=\sum_{i=1}^{N}\sqrt{(x_{\pi(i)}-x_{\pi((i+1)\%N)})^2+(y_{\pi(i)}-y_{\pi((i+1)\%N)})^2},
\end{equation}
where $\%$ denotes the modulo operator.

\section{Implementation}\label{architecture}
As we mentioned in Sec.~\ref{sec:vca}, our VCA implementation uses RNNs as a parameterized probability distribution. With RNNs, a solution $\boldsymbol{X}$ is built by the sequential realization of its elements $x_i$. In other words, solutions are sampled according to the probability chain rule
\begin{equation}\label{eq:chain_rule}
    P(\boldsymbol{X}) = P(x_1)P(x_2|x_1)\dots P(x_N |x_{N - 1}, \dots , x_2, x_1),
\end{equation}
where the product of conditional probabilities $P(x_i|x_{<i})$ provides the joint probability of the solution $\boldsymbol{X}$. Here the nature of $x_i$ depends on the optimization problem. For NSP and Max-Cut, there are $N$ binary variables $x_i\in \{0,1\}$, whereas for TSP, we have $N$ variables $x_i\in \{1, \dots, N\}$ satisfying the condition $x_i\neq x_{i'}$ for $i \neq i'$.

\begin{figure}
    \centering
    \includegraphics[width = \textwidth]{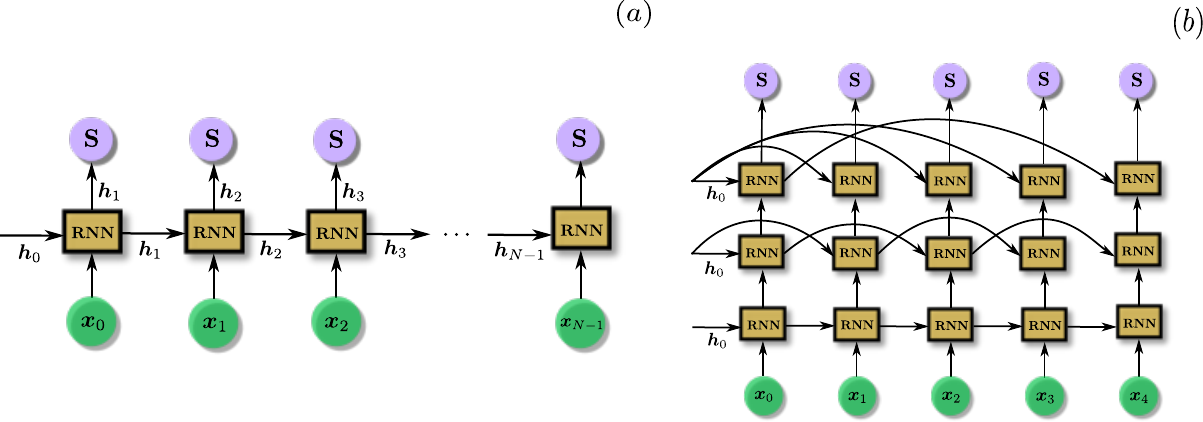}
    \caption{\textbf{The two variants of RNNs used in our study.} (a) We show  a Vanilla RNN where each RNN cell receives an input $\boldsymbol{x}_{n-1}$ and a hidden state $\boldsymbol{h}_{n-1}$ and outputs a new hidden state $\boldsymbol{h}_n$. The latter is fed to a Softmax layer (denoted by S) to output the conditional probability $\boldsymbol{P}_n$. (b) We illustrate a Dilated RNN architecture with $\lceil \log_2(N) \rceil$ layers where $N$ is the system size. This RNN uses longer recurrent connections to take care of long-range interactions.}
    \label{fig:RNN}
\end{figure}

The VCA architecture can be visualized as an $N$-sized chain of RNN cells as illustrated in Fig. \ref{fig:RNN}(a). We use the term 'site' to refer to any particular position (among $N$) in this chain. First of all, we look at the case where all RNN cells share the same set of parameters $\boldsymbol{\lambda}$  \cite{goodfellow2016deep}. We first start with the description of a Vanilla RNN cell~\cite{RNNreview}. Here, our formulation of the hidden state $\boldsymbol{h}_n$ and the conditional probability vector $\boldsymbol{P}_n$ at site $n$ are given by
\begin{gather}
    \boldsymbol{h}_n = \texttt{ELU}(W\boldsymbol{h}_{n-1} + U\boldsymbol{x}_{n-1} + \boldsymbol{b}), \label{eq:hn_vanilla}\\
    \boldsymbol{P}_n = \texttt{Softmax}(V\boldsymbol{h}_n + \boldsymbol{c}). \label{eq:pn_vanilla}
\end{gather}
In the above equations, weights $W$, $U$, $V$, and bias vectors $\boldsymbol{b}$, $\boldsymbol{c}$ jointly parameterize the RNN cells. In Eq. (\ref{eq:hn_vanilla}), $\boldsymbol{x}_{n-1}$ is the one-hot encoded vector of the variable $x_{n-1}$ sampled by the previous RNN cell. $\boldsymbol{h}_{n-1}$ is the hidden state from the previous RNN cell. The size of $\boldsymbol{h}_{n}$ is a hyperparameter called the number of memory (or hidden) units denoted as $d_h$. The latter controls the expressivity of the Vanilla RNN cell. Finally, our choice of non-linear activation function is \texttt{ELU} - exponential linear unit \cite{clevert2015fast}. In Eq. (\ref{eq:pn_vanilla}), the Softmax function, given by $\texttt{Softmax}(\sigma_i) = e^{\sigma_i}/\sum_{j=1}^{n}e^{\sigma_j}$, computes the categorical distribution from which we sample $x_n$. In this case, the conditional probability of $x_n$, denoted as $P_{\boldsymbol{\lambda}}(x_n|x_{<n})$, can be computed as 
\begin{equation}\label{eq:condprob_vanilla}
    P_{\boldsymbol{\lambda}}(x_n|x_{<n}) = \boldsymbol{P}_n \cdot \boldsymbol{x}_n,
\end{equation}
where $\cdot$ denotes the dot product operation. The set of $N$ conditional probabilities $\{P_{\boldsymbol{\lambda}}(x_1), P_{\boldsymbol{\lambda}}(x_2|x_1), \dots,  P_{\boldsymbol{\lambda}}(x_N|x_{<N})\}$ is used to compute $P_{\boldsymbol{\lambda}}(\boldsymbol{X})$ in Eq. (\ref{eq:chain_rule}). 

It should be noted that random interactions between the variables $x_i$ are best captured using RNNs with site-dependent parameters, i.e., each RNN cell has its own set of parameters dedicated to that particular site. This observation is motivated from Ref.~\cite{hibat2021variational} in which  disordered systems, such as the Sherrington-Kirkpatrick~\cite{sherrington1975solvable} and Edwards-Anderson~\cite{edwards1975theory} spin glass models, are optimized using site-dependent RNNs. In this case, we have weights $\{W_n\}_{n=1}^N$, $\{U_n\}_{n=1}^N$, $\{V_n\}_{n=1}^N$ and biases $\{\boldsymbol{b}_n\}_{n=1}^N$, $\{\boldsymbol{c}_n\}_{n=1}^N$ that are used in a Vanilla RNN architecture as follows:
\begin{gather}
    \boldsymbol{h}_n = \texttt{ELU}(W_n\boldsymbol{h}_{n-1} + U_n\boldsymbol{x}_{n-1} + \boldsymbol{b}_n)\label{eq:hn_site}
    \\
    \boldsymbol{P}_n = \texttt{Softmax}(V_n\boldsymbol{h}_n + \boldsymbol{c}_n)\label{eq:condprob_site}
\end{gather}

We now focus our attention on another RNN architecture to tackle optimization problems with long-range interactions called the Dilated RNN \cite{chang2017dilated}. This architecture uses recurrent skip connections that propagate information from sites that are far from each other as shown in Fig.~\ref{fig:RNN}(b). From the illustration, we also see that this architecture has a stack of Dilated layers that dictate the length of these skip connections given by $2^{l-1}$ where $l$ is the layer number such that $1 \leq l \leq L$. For a system of size $N$, we choose $L=\lceil \log_2(N)\rceil$ number of layers by taking inspiration from tensor networks~\cite{hibat2021variational, Vidal_2008}. Here, we index the RNN weights and biases according to the layer $l$ and site $n$ where the RNN cell is located. Thus, the RNN formulation becomes
\begin{gather}
    \boldsymbol{h}_n^{[l]} = \texttt{ELU}(W_n^{[l]}\boldsymbol{h}_{\text{max}(0,n-2^{l-1})}^{[l]} + U_n^{[l]} \boldsymbol{h}_{n}^{[l-1]} + \boldsymbol{b}_n^{[l]})\label{eq:hn_dilated}\\
    \boldsymbol{P}_n = \texttt{Softmax}(V_n\boldsymbol{h}_n^{[L]} + \boldsymbol{c}_n),\label{eq:condprob_dilated} 
\end{gather}
such that $\bm{h}_n^{[0]} = \bm{x}_{n-1}$ and $\bm{h}^{[l]}_0 = \bm{0}$ for $ 1 \leq n \leq N$ and $1 \leq l \leq L$.

We note that for the Max-Cut and NSP, we use Vanilla and Dilated RNN with site-dependent parameters to reflect the randomness in the coupling between the discrete variables. Furthermore, given the binary nature of the discrete variables for NSP and Max-Cut, the output probability vector $\boldsymbol{P}_n$ has size $2$. For TSP which enjoys the translation invariance property, we use parameters sharing for the Dilated RNN, i.e., $W_n^{[l]} = W^{[l]}$, $U_n^{[l]} = U^{[l]}$, $V_n = V$, $\boldsymbol{b}_n^{[l]} = \boldsymbol{b}^{[l]}$, $\boldsymbol{c}_n = \boldsymbol{c}$. Additionally, the output probability $\boldsymbol{P}_n$ has size $N$, such that $N$ is the number of cities.

To enforce the constraint of not visiting a city twice in TSP, we apply masking during the process of autoregressive sampling as clarified in Appendix.~\ref{sec:maskingTSP}. This trick allows us to avoid the use of an energy penalty to enforce this constraint~\cite{doi:10.7566/JPSJ.90.114002}. The masking also reduces the combinatorial search space. We remark that we conducted experiments where we replace the vanilla RNN cell with LSTM and GRU cells respectively. Compared to the models with vanilla RNN, the GRU and LSTM models do not generally improve on our results. We also note that the use of RNN architectures in our study allows us to go beyond the system sizes that can be reached by Transformer architectures~\cite{kool2018attention,bresson2021transformer}. In particular, we demonstrate that we can go beyond $N>100$ cities. This advantage is a result of the cheaper computational cost of Dilated RNNs as opposed to traditional Transformer architectures.

\section{Experiments and Discussion}\label{sec:experiments}
We now focus our attention on our experiments. 
Here we compare VCA and SA on the Max-Cut, NSP and TSP optimization problems. For all the problems, we measure the average performance of the two algorithms using the residual energy per site given by $\epsilon_{\text{res}}\equiv (\overline{H}(\boldsymbol{X})-H(\boldsymbol{X^*}))/N$. This measure quantifies the average error made on the estimation of the ground state energy $H(\boldsymbol{X^*})$. Here, $N$ denotes the system size of the problem~\cite{doi:10.1126/science.1068774}. For VCA, $\overline{H}(\boldsymbol{X})$ is estimated by taking an average over the $N_{\text{samples}}$ autoregressive samples generated from the RNNs after training. For SA, $\overline{H}(\boldsymbol{X})$ is approximated by taking the mean over different SA runs. An important observation to highlight is the ability of RNNs to sample a relatively large number of solutions at the end of annealing compared to SA, which does not offer this possibility. We also consider the best case scenario by computing the minimal residual energy per site defined as $\epsilon^{\text{min}}_{\text{res}} \equiv (H_{\text{min}}-H(\bm{X}^{*}))/N$ where $H_{\text{min}}$ is the lowest energy obtained by either SA or VCA across the different samples. To make the connection with the literature, we hightlight that $\epsilon^{\text{min}}_{\text{res}}$ is directly proportional to the optimality gap defined as $g \equiv \left (H_{\text{min}}/ H(\boldsymbol{X^*}) \right ) - 1 $~\cite{kool2018attention}. More details about the VCA hyperparameters and the SA implementation can be found in Appendix~\ref{sec:hyperparams}. 

To compare VCA and SA, $\epsilon_{\text{res}}$ is plotted against the number of annealing steps $N_{\text{annealing}}$ (see Sec.~\ref{sec:vca}). In terms of hardware, we use an Nvidia Titan XP GPU for VCA and a CPU for SA.

\begin{figure}[!htb]
    \centering
    \includegraphics[width = \textwidth]{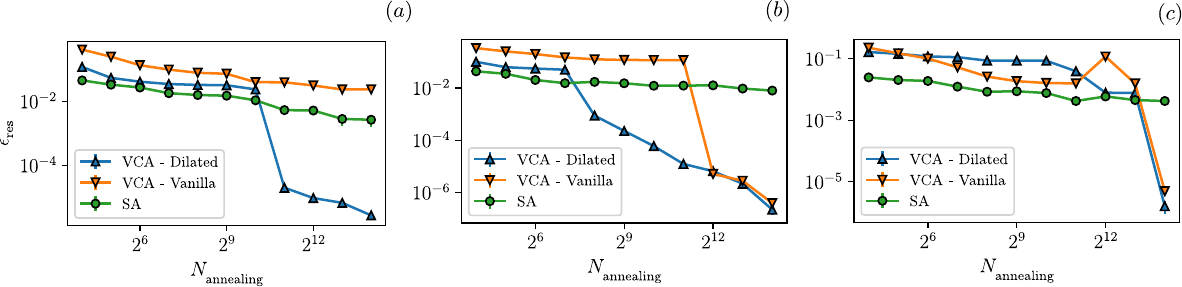}
        \caption{\textbf{Plot of the residual energy per site $\epsilon_{\text{res}}$ against the number of annealing steps $N_{\text{annealing}}$}. For all the RNN variations of VCA, we compare with SA on a Max-Cut problem with system size $N = 128$ with edge density (a) $\rho = 0.12$, (b) $\rho = 0.25$, and (c) $\rho = 0.5$. We observe that VCA outperforms SA in the limit of large $N_{\text{annealing}}$.}
        \label{fig:Maxcut128}
\end{figure}
\textbf{Max-Cut.}
We use unweighted Max-Cut instances of $N=128$ vertices and edge densities ~$\rho=0.12, 0.25, 0.5$ generated by the Rudy graph generator by G. Rinaldi~\cite{rudy}. These graphs were originally generated to benchmark the classical-quantum optimization (CQO) technique \cite{gomes2019classical}. The choice of increasing densities is used to investigate the effects of adding more complexity to the Max-Cut problem while keeping the number of vertices constant. We approximate the ground state of these graphs using the ConicBundle package by C. Helmberg from the Biq Mac Solver server~\cite{biqmac}.

We use the Vanilla RNN (see Eq. (\ref{eq:hn_site})) and Dilated RNN (see Eq. (\ref{eq:hn_dilated})) with site-dependent parameters to run VCA on this problem. As illustrated in Fig. \ref{fig:Maxcut128}, the general trend we see for both VCA and SA is that $\epsilon_{\text{res}}$ decreases with the number of annealing steps. VCA with Dilated RNNs (VCA-Dilated) outperforms SA on average starting from annealing steps $2^{11}, 2^{8}, 2^{14}$ respectively for the densities $0.12, 0.25, 0.5$. Although SA has better energies on average in the fast regime, VCA provides a better convergence at larger number of annealing steps.  This observation suggests that VCA requires a threshold number of annealing steps before it consistently converges to the ground state.

If we consider the best solutions obtained by SA and VCA (see Tab. \ref{tab:results}), we observe that for all three densities, SA finds the ground state at $N_{\text{annealing}}=2^4$. For both VCA variants, it is required to have a longer annealing time to find the ground states. The latter means that SA can find the optimal solution of our Max-Cut instances in fewer number of annealing steps compared to VCA.

\begin{figure}
    \centering
        \includegraphics[width=\textwidth]{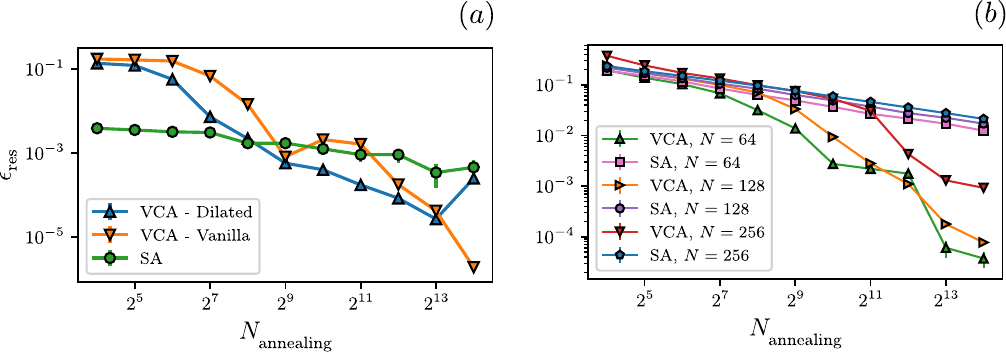}
         \caption{\textbf{Plots of the residual energy per site $\epsilon_{\text{res}}$ vs $N_{\rm annealing}$.} (a) For NSP with $15$ days and $7$ nurses, both the Vanilla and Dilated RNN variants of VCA are plotted alonside SA. (b) For TSP with system sizes $N=64, 128, 256$, VCA with Dilated RNNs and with shared parameters is plotted alongside SA. For panels (a) and (b), we see that VCA has a lower $\epsilon_{\text{res}}$ compared to SA for large $N_{\rm annealing}$.}
        \label{fig:NSP_TSP}
\end{figure}
\textbf{NSP.}
We use a configuration of $15$ days and $7$ nurses, giving us a system size of $15 \times 7 = 105$. Taking inspiration from Ref.~\cite{ikeda2019application}, we use the following NSP parameters described in Sec.~\ref{sec:nsp}: $W(d)=1, F(n)= \lfloor D/N \rfloor, \alpha=3.5, \lambda = 1.3, \gamma = 0.3$. Based on the chosen parameters, the ground state energy is $H(\boldsymbol{X}^*)=\gamma$. The proof is provided in Appendix.~\ref{sec:NSP_appendix}. 

Similar to the Max-Cut problem, we use Vanilla and Dilated RNNs with site-dependent parameters to tackle this problem. The results are presented in Fig. \ref{fig:NSP_TSP}(a). Again, we see that increasing the number of annealing steps allows VCA to reduce the average error $\epsilon_{\text{res}}$. Both VCA-Dilated and VCA-Vanilla reach lower error margins compared to SA at large $N_{\text{annealing}}$. Between annealing steps $2^{9}$ and $2^{13}$, VCA-Dilated maintains the lowest average error among all the models. For the slowest annealing schedule, $N_{\text{annealing}}=2^{14}$, VCA-Vanilla reaches the overall lowest average error, whereas VCA-Dilated sees a sudden increases in $\epsilon_{\text{res}}$. A possible reason for this sudden increase can be related to the choice of our hyperparameters, which can be further tuned. We also note that for a small number of annealing steps, SA outperforms VCA on the average case. 

Lastly, by considering the best solutions, SA finds the ground state at $N_{\text{annealing}}=2^{4}$ whereas VCA lands at the optimal solution around $N_{\text{annealing}}=2^{6}$ as illustrated in Tab.~\ref{tab:results}. The latter means that SA, similarly to Max-Cut, needs fewer $N_{\text{annealing}}$ steps compared to VCA to find optimal solutions.

\textbf{TSP.}
We use three instances of TSP with varying system sizes $N = 64, 128, 256$. Each component of the coordinates $(x,y)$ of the cities has been uniformly sampled from the range $(0,1)$. The approximation of the ground state tours of these TSP configurations are provided by the Concorde algorithm~\cite{Applegate2001} through the NEOS Server~\cite{neoserver}.

Unlike the previous problems, we use VCA with Dilated RNNs that have shared parameters owing to the translation invariance property of TSP. The focus on Dilated RNNs for this problem is motivated by the presence of long-range interactions in TSP, as well as by the overall advantage of Dilated RNNs over Vanilla RNNs for Max-Cut and NSP. The results from our experiments are presented in Fig. \ref{fig:NSP_TSP}(b). For $N=64,128$, VCA reaches a residual energy per site $\epsilon_{\text{res}} < 10^{-4}$ whereas for $N=256$, we reach an average error $\epsilon_{\text{res}} \sim 10^{-3}$. We also observe that SA reaches $\epsilon_{\text{res}} \sim 10^{-2}$ in the slow annealing regime. Overall, in the range of medium to slow annealing ($N_{\text{annealing}} \geq 2^7$), VCA demonstrates a better average performance compared to SA. Furthermore, for the largest system size, VCA requires a relatively larger $N_{\text{annealing}}$ to noticeably improve over SA highlighting the increase in complexity that arises with larger system sizes.

By considering the best case scenario, VCA finds the best tours for all three problem sizes. It also finds the exact ground state for $N = 64$, as illustrated in Tab.~\ref{tab:results}. In contrast, SA finds tours with a higher cost and gets stuck in local minima. Interestingly, this result is different from what we observed for the Max-Cut and the NSP. Thus, we conclude that if a user is interested in finding the best solutions to a particular optimization problem, VCA is a valuable alternative when SA fails to find an optimal solution.

\begin{table}[]
\scriptsize
    \centering
    \begin{tabular}{| c c c c c c c c|}
        \hline
        & & & SA & & & VCA & \\
        & Parameter &  $\epsilon^{\text{min}}_{\text{res}}$  & $N_{\text{annealing}}^*$ & Time & $\epsilon^{\text{min}}_{\text{res}}$ & $N_{\text{annealing}}^*$ & Time \\
        \hline
        NSP & 15D, 7N  & $\bm{0.0}$ & $\bm{2^{4}}$ & \bm{$3$} \textbf{s} & D: $\bm{0.0}$  & $2^{6}$ & $4 \ \text{min} \ 5 $ s \\ & & & & & V: $\bm{0.0}$ & $2^{7}$ & $1 \ \text{min} \ 25 $ s \\
        \hline
        & $\rho=0.12$ & $\bm{0.0}$ & $\bm{2^{4}}$ & \bm{$6$} \textbf{s} & D: $\bm{0.0}$  & $2^{11}$ & $45 \ \text{min} \ 32 $ s \\ & & & & & V: $\bm{0.0}$& $2^{13}$ & $1 \ \text{hr} \ 23 $ min
        \\ 
        Max-Cut & $\rho=0.25$ & $\bm{0.0}$ & $\bm{2^{4}}$ & \bm{$6$} \textbf{s} & D: $\bm{0.0}$  & $2^{8}$ & $9 \ \text{min} \ 14 $ s \\ & & & & & V: $\bm{0.0}$ & $2^{12}$ & $42 \ \text{min} \ 14 $ s \\
        & $\rho=0.50$ & $\bm{0.0}$ & $\bm{2^{4}}$ & \bm{$6$} \textbf{s} & D: $\bm{0.0}$  & $2^{14}$ & $5 \ \text{hrs} \ 35 $ min \\ & & & & & V: $\bm{0.0}$& $2^{14}$ & $2 \ \text{hrs} \ 43 $ min \\
        \hline
        & $N=64$ & $3.13 \times 10^{-3}$ & $2^{14}$ & $1 \ \text{min} \ 30 $ s & D: $\bm{0.0}$  & \bm{$2^{13}$} & $\bm{49}$ \textbf{min} $\bm{15} $ \textbf{s}\\
        TSP & $N=128$ & $9.24 \times 10^{-3}$ & $2^{14}$ & $6 \ \text{min} \ 19 $ s & D: \bm{$3.84 \times 10^{-5}$} & \bm{$2^{13}$} & $\bm{1}$ \textbf{hr} $\bm{24}$ \textbf{min}\\
        & $N=256$ & $1.75 \times 10^{-2}$ & \bm{$2^{14}$} & $25 \ \text{min} \ 42 $ s & D: \bm{$8.60 \times 10^{-4}$} & \bm{$2^{14}$} & $\bm{6}$ \textbf{hrs} $\bm{51}$ \textbf{min}\\
        \hline
    \end{tabular}
    \vspace{3mm}
    \caption{\textbf{A summary table of the best performances of VCA and SA on NSP, Max-Cut and TSP.} Here we define the minimal residual energy per site $\epsilon^{\text{min}}_{\text{res}} \equiv (H_{\text{min}}-H(\bm{X}^{*}))/N$ where $H_{\text{min}}$ is the lowest energy obtained by either SA or VCA across the different samples. `D' stands for the Dilated RNN results and `V' stands for the Vanilla RNN results. Values in bold font correspond to the lowest $N_{\text{annealing}}^*$ to find the exact or the lowest approximation to the ground state after comparing SA and VCA. Bold values also highlight the lowest $\epsilon^{\text{min}}_{\text{res}}$ as well as the lowest estimated time to find the exact or the best solution.}
    \label{tab:results}
\end{table}
\section{Conclusion}

In this work, we supplement recurrent neural networks (RNNs) with the principle of annealing under a framework called variational classical annealing (VCA) to solve real-world optimization problems. In our experiments, we find that VCA outperforms traditional simulated annealing on average in the limit of slow annealing. The latter suggests that a variational simulation of classical annealing is a great alternative for solving optimization problems. When considering the best case scenario, we conclude that VCA is best to use when SA fails to provide a good solution. The failure of SA is likely to occur for models with a glassy landscape, that has a large number of local minima, such as in spin glass models~\cite{WPE, Nokura_1987, Tayarani2014}. We also highlight that we reach large system sizes up to $256$ cities for the TSP, by virtue of the cheaper cost of RNNs compared to ordinary Transformers~\cite{TransformerPaper}. 

In our results, we see that even though VCA outperforms SA for all three problems on average in the limit of slow annealing, VCA takes more time per iteration compared to SA for the same number of annealing steps (see Tab.~\ref{tab:results}). This limitation could be mitigated by the use of meta-learning and transfer learning techniques to predict the solutions to optimization problems that are similar to each other within a shorter amount of time~\cite{MetaVMC}. Furthermore, with more hardware advancements, VCA is expected to show speed-up in the running time. We have also seen instabilities that occurred during the training of RNNs and we mitigated some of them by careful choice of the hyperparameters such as tuning the learning rate and varying the number of training samples. Overall, there is still significant space for exploration in terms of hyperparameter tuning, hardware efficiency, and choice of architectures.

From a broader perspective, we believe there are various  directions of research that could be undertaken in the future. It would be interesting to benchmark VCA against other Monte Carlo optimization algorithms such as parallel tempering~\cite{earl2005parallel}. It would be also interesting to extend our work for optimization problems with continuous variables~\cite{Fan2021} using models that can handle continuous variables~\cite{Flows2019, pmlr-v37-rezende15}. Additionally, one could also incorporate graph autoregressive networks in the VCA scheme to take the graph structure of some optimization problems into consideration~\cite{GraphRNN, Schuetz_2022}. There is also flexibility in tuning the temperature cooling schedules to potentially improve the performance of VCA~\cite{Mills_2020}. We believe that the combination of deep learning techniques and the principle of annealing is an interesting direction of research that could potentially improve our current solutions to optimization problems.

\section*{Acknowledgments}
We are grateful to Juan Carrasquilla for providing valuable insights on the manuscript. We also thank Amine Mohamed Aboussalah, Estelle Inack, Jack Raymond, Rochisha Agarwal for valuable discussions. We would like to acknowledge Wasif Ahmed for making the connection between the authors of this paper. We are also grateful for Vector Institute compute resources that made our numerical simulations possible. M.H. acknowledges support from Mitacs through the Mitacs Accelerate program.

\section*{Reproducibility Code}
The code we use to produce our results can be found in Ref.~\cite{github_code}.

\bibliographystyle{unsrt}
\bibliography{bibliography.bib}

\begin{thebibliography}{10}

\bibitem{hibat2021variational}
Mohamed Hibat-Allah, Estelle~M Inack, Roeland Wiersema, Roger~G Melko, and Juan
  Carrasquilla.
\newblock Variational neural annealing.
\newblock {\em Nature Machine Intelligence}, 3(11):952--961, 2021.

\bibitem{Lucas2014}
Andrew Lucas.
\newblock Ising formulations of many np problems.
\newblock {\em Frontiers in Physics}, 2, 2014.

\bibitem{SApaper}
S.~Kirkpatrick, C.~D. Gelatt, and M.~P. Vecchi.
\newblock Optimization by simulated annealing.
\newblock {\em Science}, 220(4598):671--680, 1983.

\bibitem{bresson2021transformer}
Xavier Bresson and Thomas Laurent.
\newblock The transformer network for the traveling salesman problem.
\newblock {\em arXiv preprint arXiv:2103.03012}, 2021.

\bibitem{gomes2019classical}
Joseph Gomes, Keri~A McKiernan, Peter Eastman, and Vijay~S Pande.
\newblock Classical quantum optimization with neural network quantum states.
\newblock {\em arXiv preprint arXiv:1910.10675}, 2019.

\bibitem{carleo2017solving}
Giuseppe Carleo and Matthias Troyer.
\newblock Solving the quantum many-body problem with artificial neural
  networks.
\newblock {\em Science}, 355(6325):602--606, 2017.

\bibitem{Sinchenko2019}
Semyon Sinchenko and Dmitry Bazhanov.
\newblock The deep learning and statistical physics applications to the
  problems of combinatorial optimization, 2019.

\bibitem{zhao2020natural}
Tianchen Zhao, Giuseppe Carleo, James Stokes, and Shravan Veerapaneni.
\newblock Natural evolution strategies and variational monte carlo.
\newblock {\em Machine Learning: Science and Technology}, 2(2):02LT01, 2020.

\bibitem{sorella1998green}
Sandro Sorella.
\newblock Green function monte carlo with stochastic reconfiguration.
\newblock {\em Physical review letters}, 80(20):4558, 1998.

\bibitem{RNNAnnealing}
Mohamed Hibat-Allah, Roger~G Melko, and Juan Carrasquilla.
\newblock Supplementing recurrent neural network wave functions with symmetry
  and annealing to improve accuracy.
\newblock {\em Machine Learning and the Physical Sciences (NeurIPS 2021)},
  2021.

\bibitem{vaswani2017attention}
Ashish Vaswani, Noam Shazeer, Niki Parmar, Jakob Uszkoreit, Llion Jones,
  Aidan~N Gomez, {\L}ukasz Kaiser, and Illia Polosukhin.
\newblock Attention is all you need.
\newblock {\em Advances in neural information processing systems}, 30, 2017.

\bibitem{kool2018attention}
Wouter Kool, Herke Van~Hoof, and Max Welling.
\newblock Attention, learn to solve routing problems!
\newblock {\em arXiv preprint arXiv:1803.08475}, 2018.

\bibitem{deudon2018learning}
Michel Deudon, Pierre Cournut, Alexandre Lacoste, Yossiri Adulyasak, and
  Louis-Martin Rousseau.
\newblock Learning heuristics for the tsp by policy gradient.
\newblock In {\em International conference on the integration of constraint
  programming, artificial intelligence, and operations research}, pages
  170--181. Springer, 2018.

\bibitem{bayeslearningrule2020}
Xiangming Meng, Roman Bachmann, and Mohammad~Emtiyaz Khan.
\newblock Training binary neural networks using the bayesian learning rule,
  2020.

\bibitem{SAC2018}
Tuomas Haarnoja, Aurick Zhou, Pieter Abbeel, and Sergey Levine.
\newblock Soft actor-critic: Off-policy maximum entropy deep reinforcement
  learning with a stochastic actor, 2018.

\bibitem{VAEannealing}
Hao Fu, Chunyuan Li, Xiaodong Liu, Jianfeng Gao, Asli Celikyilmaz, and Lawrence
  Carin.
\newblock Cyclical annealing schedule: A simple approach to mitigating kl
  vanishing, 2019.

\bibitem{curriculumlearning}
Michal Lisicki, Arash Afkanpour, and Graham~W. Taylor.
\newblock Evaluating curriculum learning strategies in neural combinatorial
  optimization, 2020.

\bibitem{BENGIO2021405}
Yoshua Bengio, Andrea Lodi, and Antoine Prouvost.
\newblock Machine learning for combinatorial optimization: A methodological
  tour d’horizon.
\newblock {\em European Journal of Operational Research}, 290(2):405--421,
  2021.

\bibitem{wu2019solving}
Dian Wu, Lei Wang, and Pan Zhang.
\newblock Solving statistical mechanics using variational autoregressive
  networks.
\newblock {\em Physical Review Letters}, 122(8), Feb 2019.

\bibitem{condmat7020038}
Estelle~M. Inack, Stewart Morawetz, and Roger~G. Melko.
\newblock Neural annealing and visualization of autoregressive neural networks
  in the newman-moore model.
\newblock {\em Condensed Matter}, 7(2), 2022.

\bibitem{osogami2000classification}
Takayuki Osogami and Hiroshi Imai.
\newblock Classification of various neighborhood operations for the nurse
  scheduling problem.
\newblock In {\em International Symposium on Algorithms and Computation}, pages
  72--83. Springer, 2000.

\bibitem{ikeda2019application}
Kazuki Ikeda, Yuma Nakamura, and Travis~S Humble.
\newblock Application of quantum annealing to nurse scheduling problem.
\newblock {\em Scientific reports}, 9(1):1--10, 2019.

\bibitem{glover2019quantum}
Fred Glover, Gary Kochenberger, and Yu~Du.
\newblock Quantum bridge analytics i: a tutorial on formulating and using qubo
  models.
\newblock {\em 4OR}, 17(4):335--371, 2019.

\bibitem{karp1972reducibility}
Richard~M Karp.
\newblock Reducibility among combinatorial problems.
\newblock In {\em Complexity of computer computations}, pages 85--103.
  Springer, 1972.

\bibitem{papadimitriou1977euclidean}
Christos~H Papadimitriou.
\newblock The euclidean travelling salesman problem is np-complete.
\newblock {\em Theoretical computer science}, 4(3):237--244, 1977.

\bibitem{goodfellow2016deep}
Ian Goodfellow, Yoshua Bengio, and Aaron Courville.
\newblock {\em Deep learning}.
\newblock MIT press, 2016.

\bibitem{RNNreview}
Zachary~C. Lipton, John Berkowitz, and Charles Elkan.
\newblock A critical review of recurrent neural networks for sequence learning,
  2015.

\bibitem{clevert2015fast}
Djork-Arn{\'e} Clevert, Thomas Unterthiner, and Sepp Hochreiter.
\newblock Fast and accurate deep network learning by exponential linear units
  (elus).
\newblock {\em arXiv preprint arXiv:1511.07289}, 2015.

\bibitem{sherrington1975solvable}
David Sherrington and Scott Kirkpatrick.
\newblock Solvable model of a spin-glass.
\newblock {\em Physical review letters}, 35(26):1792, 1975.

\bibitem{edwards1975theory}
Samuel~Frederick Edwards and Phil~W Anderson.
\newblock Theory of spin glasses.
\newblock {\em Journal of Physics F: Metal Physics}, 5(5):965, 1975.

\bibitem{chang2017dilated}
Shiyu Chang, Yang Zhang, Wei Han, Mo~Yu, Xiaoxiao Guo, Wei Tan, Xiaodong Cui,
  Michael Witbrock, Mark Hasegawa-Johnson, and Thomas~S Huang.
\newblock Dilated recurrent neural networks.
\newblock {\em arXiv preprint arXiv:1710.02224}, 2017.

\bibitem{Vidal_2008}
G.~Vidal.
\newblock Class of quantum many-body states that can be efficiently simulated.
\newblock {\em Physical Review Letters}, 101(11), sep 2008.

\bibitem{doi:10.7566/JPSJ.90.114002}
Vladimir Vargas-Calderón, Nicolas Parra-A., Herbert Vinck-Posada, and Fabio~A.
  González.
\newblock Many-qudit representation for the travelling salesman problem
  optimisation.
\newblock {\em Journal of the Physical Society of Japan}, 90(11):114002, 2021.

\bibitem{doi:10.1126/science.1068774}
Giuseppe~E. Santoro, Roman Martoňák, Erio Tosatti, and Roberto Car.
\newblock Theory of quantum annealing of an ising spin glass.
\newblock {\em Science}, 295(5564):2427--2430, 2002.

\bibitem{rudy}
\url{http://www-user.tu-chemnitz.de/~helmberg/rudy.tar.gz}.

\bibitem{biqmac}
Biq~Mac Solver.
\newblock \url{https://biqmac.aau.at/}.

\bibitem{Applegate2001}
David Applegate, Robert Bixby, Va{\v{s}}ek Chv{\'a}tal, and William Cook.
\newblock {\em TSP Cuts Which Do Not Conform to the Template Paradigm}, pages
  261--303.
\newblock Springer Berlin Heidelberg, Berlin, Heidelberg, 2001.

\bibitem{neoserver}
\url{https://neos-server.org/neos/solvers/co:concorde/TSP.html}.

\bibitem{WPE}
Firas Hamze, Jack Raymond, Christopher~A. Pattison, Katja Biswas, and Helmut~G.
  Katzgraber.
\newblock Wishart planted ensemble: A tunably rugged pairwise ising model with
  a first-order phase transition.
\newblock {\em Phys. Rev. E}, 101:052102, May 2020.

\bibitem{Nokura_1987}
K~Nokura.
\newblock An heuristic approach to the structure of local minima of the
  sherrington-kirkpatrick model.
\newblock {\em Journal of Physics A: Mathematical and General},
  20(17):L1203--L1205, dec 1987.

\bibitem{Tayarani2014}
Mohammad-H. Tayarani-N. and Adam Prügel-Bennett.
\newblock On the landscape of combinatorial optimization problems.
\newblock {\em IEEE Transactions on Evolutionary Computation}, 18(3):420--434,
  2014.

\bibitem{TransformerPaper}
Ashish Vaswani, Noam Shazeer, Niki Parmar, Jakob Uszkoreit, Llion Jones,
  Aidan~N. Gomez, Lukasz Kaiser, and Illia Polosukhin.
\newblock Attention is all you need, 2017.

\bibitem{MetaVMC}
Tianchen Zhao, James Stokes, Oliver Knitter, Brian Chen, and Shravan
  Veerapaneni.
\newblock Meta variational monte carlo, 2020.

\bibitem{earl2005parallel}
David~J Earl and Michael~W Deem.
\newblock Parallel tempering: Theory, applications, and new perspectives.
\newblock {\em Physical Chemistry Chemical Physics}, 7(23):3910--3916, 2005.

\bibitem{Fan2021}
Qingsong Fan, Haisong Huang, Qipeng Chen, Liguo Yao, Kai Yang, and Dong Huang.
\newblock A modified self-adaptive marine predators algorithm: Framework and
  engineering applications.
\newblock {\em Eng. with Comput.}, 38(4):3269–3294, aug 2022.

\bibitem{Flows2019}
George Papamakarios, Eric Nalisnick, Danilo~Jimenez Rezende, Shakir Mohamed,
  and Balaji Lakshminarayanan.
\newblock Normalizing flows for probabilistic modeling and inference.
\newblock 2019.

\bibitem{pmlr-v37-rezende15}
Danilo Rezende and Shakir Mohamed.
\newblock Variational inference with normalizing flows.
\newblock In Francis Bach and David Blei, editors, {\em Proceedings of the 32nd
  International Conference on Machine Learning}, volume~37 of {\em Proceedings
  of Machine Learning Research}, pages 1530--1538, Lille, France, 07--09 Jul
  2015. PMLR.

\bibitem{GraphRNN}
Jiaxuan You, Rex Ying, Xiang Ren, William~L. Hamilton, and Jure Leskovec.
\newblock Graphrnn: Generating realistic graphs with deep auto-regressive
  models, 2018.

\bibitem{Schuetz_2022}
Martin J.~A. Schuetz, J.~Kyle Brubaker, and Helmut~G. Katzgraber.
\newblock Combinatorial optimization with physics-inspired graph neural
  networks.
\newblock {\em Nature Machine Intelligence}, 4(4):367--377, apr 2022.

\bibitem{Mills_2020}
Kyle Mills, Pooya Ronagh, and Isaac Tamblyn.
\newblock Finding the ground state of spin hamiltonians with reinforcement
  learning.
\newblock {\em Nature Machine Intelligence}, 2(9):509--517, sep 2020.

\bibitem{github_code}
Github repository.
\newblock \url{https://github.com/RNN-VCA-CO/RNN-VCA-CO}.

\bibitem{kingma2014adam}
Diederik~P Kingma and Jimmy Ba.
\newblock Adam: A method for stochastic optimization.
\newblock {\em arXiv preprint arXiv:1412.6980}, 2014.

\bibitem{Marto_k_2004}
Roman Marto{\v{n} }{\'{a}}k, Giuseppe~E. Santoro, and Erio Tosatti.
\newblock Quantum annealing of the traveling-salesman problem.
\newblock {\em Physical Review E}, 70(5), nov 2004.

\bibitem{PhysRevResearch.2.023358}
Mohamed Hibat-Allah, Martin Ganahl, Lauren~E. Hayward, Roger~G. Melko, and Juan
  Carrasquilla.
\newblock Recurrent neural network wave functions.
\newblock {\em Phys. Rev. Research}, 2:023358, Jun 2020.

\bibitem{https://doi.org/10.48550/arxiv.1611.09940}
Irwan Bello, Hieu Pham, Quoc~V. Le, Mohammad Norouzi, and Samy Bengio.
\newblock Neural combinatorial optimization with reinforcement learning, 2016.

\end{thebibliography}

\newpage
\appendix

\section{Hyperparameters and Training}
\label{sec:hyperparams}
The VCA hyperparameters are detailed in Tab.~\ref{tab:VCAhyperparams}. Our SA implementation is based on Metropolis moves. At the initial temperatures $T_0$, we perform $N_{\text{warmup}}$ warmup steps where each step corresponds to a sweep of $N$ Metropolis moves, where $N$ is the number of variables. After this step, we alternate between annealing steps (by decreasing temperature) and equilibrium steps $N_{\text{eq}}$. The latter is similar in spirit to $N_{\text{train}}$ in VCA. The hyperparameters of SA (see Tab.~\ref{tab:SAhyperparams}) are chosen to be consistent with VCA hyperparameters.

To train our RNN architectures, we minimize the variational free energy function $F_{\boldsymbol{\lambda}}(t) = \langle H(\boldsymbol{X}) \rangle_{\boldsymbol{\lambda}} - T(t)S(P_{\boldsymbol{\lambda}})$. Here we approximate the true variational free energy at any given temperature by taking the average over $N_{\text{samples}}$ solutions sampled from the RNN distribution $P_{\boldsymbol{\lambda}}$. This approximation is given by
\begin{equation}\label{eq:free_energy_approx}
    F_{\boldsymbol{\lambda}}(t) \approx \frac{1}{N_{\text{samples}}} \sum_{i=1}^{N_{\text{samples}}} \left (H(\boldsymbol{X}_i) + T(t)\log(P_{\boldsymbol{\lambda}}(\boldsymbol{X}_i)) \right ).
\end{equation}
Furthermore, the gradients of the variational free energy $\partial_{\boldsymbol{\lambda}} F_{\boldsymbol{\lambda}}(t)$ have a similar expression as shown in Ref.~\cite{hibat2021variational}. We also optimize the relevant parameters $\boldsymbol{\lambda}$ with the Adam optimizer \cite{kingma2014adam}.

\begin{table}
\begin{center}

\caption{A table of the hyperparameters used for the VCA experiments in our study.}
\label{tab:VCAhyperparams}
\vspace{3mm}
\centering
\begin{tabular}{ | m{2.5cm} | m{2.5cm}| m{2.5cm} | m{2.5cm}|} 
  \hline
  Hyperparameter & Max-Cut (Fig. \ref{fig:Maxcut128}) & NSP (Fig. \ref{fig:NSP_TSP}(a)) & TSP (Fig. \ref{fig:NSP_TSP}(b)) \\ 
  \hline
  Architecture & Dilated and Vanilla RNN  & Dilated and Vanilla RNN & Dilated RNN with weight sharing \\ 
  \hline 
  $N_{\text{samples}}$& $50$ ($5\times 10^5$ after annealing)  &$50$ ($5\times 10^5$ after annealing) &$50$ ($10^6$ after annealing) \\ 
  \hline
  $N_{\text{train}}$ &$5$ &$5$ &$5$\\ 
  \hline 
  $N_{\text{warmup}}$ &$1000$ &$1000$ &$2000$ \\ 
  \hline 
  $N_{\text{annealing}}$ &  $[2^4,2^5,....,2^{14}]$ &$[2^4,2^5,....,2^{14}]$ &$[2^4,2^5,....,2^{14}]$\\ 
  \hline 
  Learning rate & $1\times 10^{-4}$ & $5 \times 10^{-4}$ & $1 \times 10^{-3}$\\ 
  \hline 
  $T_0$ &$2.0$ &$2.0$ &$2.0$\\
  \hline 
  RNN hidden units &$40$ &$40$ &$40$\\ 
  \hline 
  Seed  &$111$ &$111$ &$111$ \\ 
  \hline
\end{tabular}
\end{center}
\end{table}

\begin{table}
\begin{center}

\caption{A table of the hyperparameters used for the SA experiments in this paper. $N_{\text{samples}}$ corresponds to the number of SA runs similarly to the batch size in VCA.}
\label{tab:SAhyperparams}
\vspace{3mm}

\centering
\begin{tabular}{ | m{2.5cm} | m{2.5cm}| m{2.5cm} | m{2.5cm}|} 
  \hline
  Hyperparameter & Max-Cut (Fig. \ref{fig:Maxcut128}) & NSP (Fig. \ref{fig:NSP_TSP}(a)) & TSP (Fig. \ref{fig:NSP_TSP}(b)) \\ 
  \hline
  Metropolis move & Bit flips  & Bit flips & Permutations (2-Opt)~\cite{Marto_k_2004} \\ 
  \hline 
  $N_{\text{samples}}$& $50$  &$50$ &$50$ \\ 
  \hline
  $N_{\text{eq}}$ &$5$ &$5$ &$5$\\ 
  \hline 
  $N_{\text{warmup}}$ &$1000$ &$1000$ &$2000$ \\ 
  \hline 
  $N_{\text{annealing}}$ &  $[2^4,2^5,....,2^{14}]$ &$[2^4,2^5,....,2^{14}]$ &$[2^4,2^5,....,2^{14}]$\\ 
  \hline 
  $T_0$ &$2.0$ &$2.0$ &$2.0$\\
  \hline 
  Seed  &$111$ &$111$ &$111$ \\ 
  \hline
\end{tabular}
\end{center}
\end{table}

\section{Masking in TSP}
\label{sec:maskingTSP}
To avoid revisiting the same cities during the processing of autoregressive sampling in VCA, we take inspiration from Ref.~\cite{PhysRevResearch.2.023358}, where constraints were applied in the context of quantum physics, and from Ref.~\cite{https://doi.org/10.48550/arxiv.1611.09940} in the context of TSP. Here we apply masking on the visited cities when computing the conditional probabilities $P_{\boldsymbol{\theta}}(c_i|c_{i-1}, \ldots, c_2, c_1)$ at the level of the Softmax layer (see Eq.~(\ref{eq:condprob_vanilla})), where $c_i \in \{1,2,\ldots, N \}$ corresponds to the city to be visited at step $i$ and ${\boldsymbol{\theta}}$ are the parameters of our model. 

The masking at step $i$ is done as follows:
\begin{itemize}
    \item If cities $(c_1, c_2, \ldots c_{i-1})$ were visited, where $c_j \in \{1,2,\ldots, N \}$, then $P_{\boldsymbol{\theta}}(c_j|c_{i-1}, \ldots, c_2, c_1)$ is set to zero for $1 \leq j \leq i-1$.
    \item After the previous step, the conditional probability $\boldsymbol{P}_{\boldsymbol{\theta}}(.|c_{i-1}, \ldots, c_2, c_1)$ is renormalized to 1.
\end{itemize}
After implementing these steps, our autoregressive model generates permutations $\pi$ of the cities. We note that this masking trick can be implemented in parallel across the batch size.

\section{NSP Ground State Energy}\label{sec:NSP_appendix}

Instead of deriving the ground state energy for NSP with days $D=15$ and nurses $N=7$, we show near identical calculations for NSP with a relatively simpler configuration of $D=5, N=2$. The NSP parameters used in our experiment are shown in Tab.~\ref{tab:nsp_params}. Looking at the ground state (see Tab.~\ref{tab:nsp_params}(a)), we see that the hard nurse constraint is respected since neither nurse $n_1$ nor nurse $n_2$ has to work two days in a row. Referring back to the penalty functions from Sec.~\ref{sec:nsp}, we obtain
\begin{equation*}
    H_1(X^*) = 0.
\end{equation*}
We also know that each of the days only requires $W(d)=1$ amount of workforce. Therefore, the hard shift constraint is also satisfied since there is only one nurse assigned each day:
\begin{equation*}
    H_2(X^*) = 0.
\end{equation*}
However, the soft nurse constraint is broken exactly once because $5\%2=1$ where $\%$ is the modulo operator. This observation leaves exactly one day (here $d_5$) to which a nurse will be assigned to (in this case $n_1$) who has to work an extra day. Therefore
\begin{gather*}
    H_3(X^*) = 1. \\
    \intertext{Summing the penalty functions with their respective scaling, we obtain}
    H(X^*) = H_1(X^*) + \lambda H_2(X^*) + \gamma H_3(X^*) = \gamma.
\end{gather*}
One can notice that it is energetically favorable to break the soft nurse constraint once than the hard shift constraint as $\gamma < \lambda, \alpha$. Thus, a solution $X'$ that does not have any nurse working on $d_5$ can not be the ground state (see Tab.~\ref{tab:nsp_params}(b)).

\begin{table}
\caption{The NSP parameters used in our experiments along with (a) a ground state solution and (b) a sub-optimal solution for NSP with $D=5, N=2$.}\label{tab:nsp_params}
\vspace{3mm}
\begin{subtable}{\textwidth}
    \centering
    \begin{tabular}{|c c c c c|}
    \hline
        $\alpha$ & $\lambda$ & $\gamma$ & $W(d)$ & $F(n)$ \\
        \hline
        $3.5$ & $1.3$ & $0.3$ & $1$ & $\lfloor D/N \rfloor$ \\
        \hline
    \end{tabular}
\end{subtable}
\bigskip
   \centering
    \subfloat[Ground state solution $X^*$.]{
     \begin{tabular}{|c c c c c c|}
        \hline
        & $d_1$ & $d_2$ & $d_3$ & $d_4$ & $d_5$ \\
        \hline
        $n_1$ & 1 & 0 & 1 & 0 & 1 \\
        \hline
        $n_2$ & 0 & 1 & 0 & 1 & 0 \\
        \hline
    \end{tabular}
    }
\hspace{5mm}
    \centering
    \subfloat[Sub-optimal solution $X'$.]{
     \begin{tabular}{|c c c c c c|}
        \hline
        & $d_1$ & $d_2$ & $d_3$ & $d_4$ & $d_5$ \\
        \hline
        $n_1$ & 1 & 0 & 1 & 0 & 0 \\
        \hline
        $n_2$ & 0 & 1 & 0 & 1 & 0 \\
        \hline
    \end{tabular}
    }
\end{table}

\end{document}